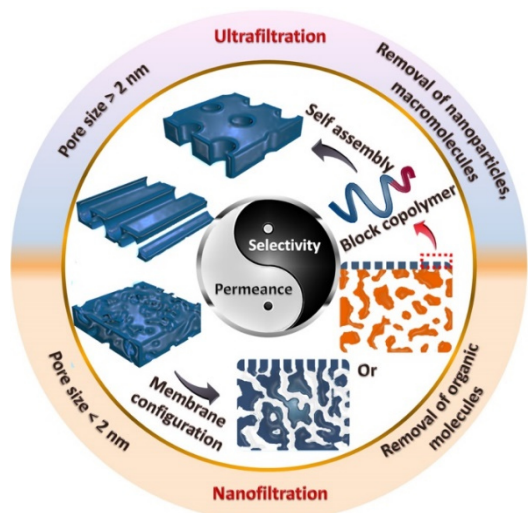

# Porous block copolymer separation membranes for 21st century sanitation and hygiene

Leiming Guo,[a, b] Yong Wang[c] and Martin Steinhart[b]

[a] College of Environmental and Chemical Engineering, Shanghai University, Shanghai, 200444, China. E-mail: leiming_guo@shu.edu.cn
[b] Institut für Chemie neuer Materialien and CellNanOs, Universität Osnabrück, Barbarastr. 7, 49076 Osnabrück, Germany. E-mail: martin.steinhart@uos.de
[c] State Key Laboratory of Materials-Oriented Chemical Engineering, College of Chemical Engineering, Nanjing Tech University, Nanjing 211816, Jiangsu, China. E-mail: yongwang@njtech.edu.cn

Removing hazardous particulate and macromolecular contaminants as well as viruses with sizes from a few nm up to the 100-nm-range from water and air is crucial for ensuring sufficient sanitation and hygiene for a growing world population. To this end, high-performance separation membranes are needed that combine high permeance, high selectivity and sufficient mechanical stability under operating conditions. However, design features of separation membranes enhancing permeance reduce selectivity and *vice versa*. Membrane configurations combining high permeance and high selectivity suffer in turn from a lack of mechanical robustness. These problems may be tackled by using block copolymers (BCPs) as a material platform for the design of separation membranes. BCPs are macromolecules that consist of two or more chemically distinct block segments, which undergo microphase separation yielding a wealth of ordered nanoscopic domain structures. Various methods allow the transformation of these nanoscopic domain structures into customized nanopore systems with pore sizes in the sub-100-nm range and with narrow pore size distributions. This tutorial review summarizes design strategies for nanoporous state-of-the-art BCP separation membranes, their preparation, their device integration and their use for water purification.

**Key learning points**
1. Comprehensive overview of the state-of-the-art regarding BCP-based separation membranes.
2. Strategies for the structural and morphological design of BCP membranes.
3. Correlation between the morphology of BCP-based membranes and separation benchmark parameters such as selectivity and permeance.
4. Device integration and use cases of BCP-based membranes.

# 1. Introduction

The supply of clean water has emerged as a major challenge for mankind. The shortage of clean water is further exacerbated by continuous population growth and by anthropogenic climate change. It is of significant importance to pay close attention to the hygiene of water since the quality of water available to humans is highly correlated to human health. The problems caused by the lack of clean water may be mitigated by the engineering of water redistribution and storage, sea water desalination and wastewater treatment.[1] Especially wastewater treatment is a demanding task since water pollution may comprise not only dissolved small molecules such as detergents, active pharmaceutical ingredients and nutrients but also pathogens and particulate pollutants such as microplastics. Sophisticated separation techniques including, for example, distillation, crystallization of pollutants, extraction, adsorption separation and membrane separation are employed to remove potentially hazardous pollutants.[2] Compared to other water purification methods, membrane separation is environmentally friendly, energy-saving and associated with a small ecologic footprint.[3] Separation membranes rejecting pollutants and solutes larger than 100 nm are called microfiltration membranes, separation membranes rejecting pollutants and solutes larger than 2 nm are called ultrafiltration membranes and separation membranes rejecting pollutants and solutes smaller than 2 nm are called nanofiltration membranes.[4] For efficient operation, separation membranes should be characterized by high permeance as well as by high selectivity. Unfortunately, structural and morphological features of separation membranes that result in high permeance often reduce selectivity and *vice versa*. To overcome this conflict known as trade-off effect, careful design of the separation membranes regarding pore size, pore geometry, tortuosity (mean length of curvilinear transport pathways through the membrane divided by the membrane thickness, cf. Fig. 1) and porosity (pore volume divided by overall membrane volume) is required. However, membrane designs potentially overcoming the trade-off effect may suffer from poor mechanical stability. Typically, only gradient membranes and composite systems containing thin size-selective layers with narrow pore size distributions meet the desired performance benchmarks.

Several materials have been studied for their use in different separation membrane architectures. Separation membranes may consist of inorganic oxides such as alumina, titania or zirconia. Nanoporous alumina membranes with narrow pore size distributions are commercially available (*e.g.*, Whatman Anodisc). However, membranes consisting of inorganic oxides that are thin enough for separation processes characterized by high permeances typically suffer from high brittleness. This drawback may be overcome by the use of polymer membranes. Currently, separation membranes consisting of commercially available synthetic organic homopolymers such as polyvinylidene difluoride (PVDF), polysulfone (PSF), polypropylene (PP), polyethersulfone (PES), polyamide (PA) and polyimide (PI) have the largest share of the separation membrane market. Synthetic organic homopolymers are much easier to process than inorganic oxides – homopolymer membranes are commonly produced by non-solvent-induced phase separation (NIPS),[5] an up-scalable industrial manufacturing process. Furthermore, homopolymer membranes exhibit reasonable mechanical properties. A main drawback of commercially available homopolymer membranes is their broad pore size distribution, which reduces the selectivity of separation processes. Homopolymer membranes with narrow pore size distributions may be obtained by the etching of ion tracks generated in homopolymer films by ion bombardment.[6] While the wall chemistry of track-etched pores can easily be customized, ion bombardment can be carried out only in specialized large research facilities, and ion-track membranes typically have a thickness exceeding several µm. However, for separation applications thinner membranes are required.[7-9] Track-etched membranes can, in principle, be prepared with high porosities but contain then high proportions of intersecting pores – resulting in insufficient mechanical strength of the membrane scaffold. As a result, commercial track-etched membranes typically have porosities lower than 3% and correspondingly low permeances.

To overcome the aforementioned problems, membrane materials with reasonable mechanical properties are needed that allow the realization of pore systems characterized by narrow pore size distributions, regular pore shapes and high porosities. Potential membrane materials including membrane proteins like aquaporins, peptides, carbon nanotubes, graphene oxides and polymerizable surfactants have been considered for the design of next-generation separation membranes for water treatment.[10] In specific application scenarios, membranes made from these materials remove contaminants from aqueous solutions with sufficient selectivity while they still exhibit high water permeances. However, only limited adaptations of the textural properties of these membranes are possible so as to meet the requirements of given use cases. Some of these materials, such as graphene oxides, are unstable in water. Sufficient stability of the corresponding membranes can only be achieved by additional measures such as grafting steps,[11] which inevitably change the lamellar spacings (pore sizes) of graphene oxide and impede its use for separation purposes.

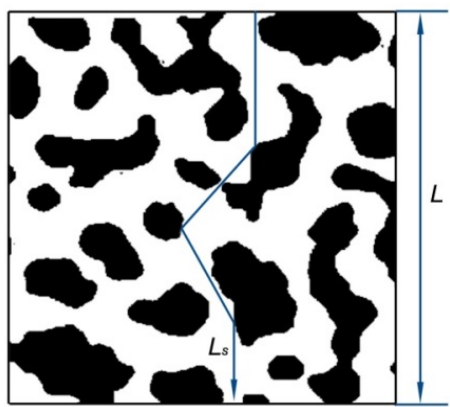

**Fig. 1.** Sketch illustrating the tortuosity $\tau = L_s / L$ of the pore system of a membrane. $L_s$ is the mean length of the curvilinear transport pathways through the membrane, $L$ is the membrane thickness.

The preparation of separation membranes from block copolymers (BCPs) enables flexible tailoring of pore sizes, pore shapes, tortuosities, porosities and the surface chemistry of the pore walls. Moreover, thin size-selective BCP layers with adequate mechanical stabilities can easily be integrated into gradient and composite membrane designs. Therefore, the looming shortage of clean water may be alleviated by the availability of advanced water treatment systems based on BCP membranes. BCP membranes may be employed as a component especially of small mobile water purification devices that can be operated where needed. BCP membranes might also be used as filters in ventilation and air conditioning systems. Several reviews have covered specific aspects of BCP membranes.[12-16] A recent review by Hampu et al. comprehensively addresses the state of the art regarding block copolymer ultrafiltration membranes.[17] In this tutorial review, we aim at the explicit introduction of basic terms and concepts relevant to the design and use of BCP ultrafiltration and nanofiltration membranes. In Section 2, key parameters to assess the performance of BCP separation membranes as well as the intertwined and often conflicting relation between selectivity and permeance are discussed. Section 3 introduces the peculiar features of BCPs as a material platform for ultrafiltration membranes. The state of the art regarding the design of BCP ultrafiltration membranes will be addressed, focussing on textural and morphological features of the pore systems of BCP membranes as well as on the preparation of BCP membranes by generation of pores in BCP films and monoliths. In section 4, strategies for the design of complex ultrafiltration membrane configurations overcoming the trade-off between permeance and selectivity are discussed. Section 5 deals with the integration of BCP ultrafiltration membranes into device configurations for water purification and corresponding use cases of BCP ultrafiltration membranes. Possible designs for BCP nanofiltration membranes and corresponding application examples are reviewed in Section 6. We conclude this tutorial review with a summary and with perspectives for the further development and future applications of BCP separation membranes.

## 2. Separation performance: key parameters

In the following, it will be discussed how the performance of separation membranes is quantified. Selectivity and permeance are introduced as key benchmark parameters, and the trade-off associated with their concurrent optimization is discussed.

### 2.1 Selectivity

The selectivity is a measure of the efficiency, with which solutes – such as contaminants – are removed from a gaseous or liquid fluid to be purified. Commonly, the selectivity is represented by the rejection $R$ and the molecular-weight-cut-off (*MWCO*). $R$ is the fraction of a solute that does not pass a separation membrane:

$$R = (1 - \frac{C_2}{C_1}) \times 100\% \qquad (1)$$

$C_1$ is the solute concentration in the fluid prior to filtering (referred to "feed"), and $C_2$ the solute concentration in the purified fluid after filtering (referred to "permeate"). The *MWCO* is defined in such a way that a membrane rejects at least 90 % of a solute with a molecular weight $M$ equal or larger than the *MWCO*. *MWCO* values are determined by using aqueous test solutions containing polymers such as polyethylene oxide (PEO)[18] or dextran[19] with various well-defined molecular weights. The effective pore diameters $d$ can then be estimated by empirically correlating the molecular weight $M$ of the largest polymer solute that passes a tested separation membrane and its hydrodynamic radius $r$, which is the radius of a hard sphere that diffuses like the solute.

### 2.2 Permeance

The permeance $P_W$ (L·h$^{-1}$·m$^{-2}$·bar$^{-1}$) is the fluid volume $V$ (L) that passes a separation membrane per time $t$ (h), per membrane area $A$ (m²) and per pressure drop $\Delta p$ (bar):

$$P_W = \frac{V}{At\Delta p} \qquad (2)$$

On the other hand, the velocity $v$ (m·s$^{-1}$) of the flowing fluid equals the fluid volume $V$ passing the separation membrane per membrane area $A$ and per time $t$. Therefore, we obtain:

$$v = \frac{V}{At} = P_W \Delta p \qquad (3)$$

Assuming that a separation membrane contains straight cylindrical pores oriented parallel to the flow direction of the fluid and assuming that the flow through the membrane pores is laminar, the effective pore diameter $d$ of the membranes can be obtained from $v$ using Hagen-Poiseuille's law:[20]

$$v = \frac{\varepsilon}{\tau^2} \left(\frac{\Delta p d^2}{32 \mu L}\right) \qquad (4)$$

Here, $\mu$ is the dynamic viscosity of the fluid (bar·s$^{-1}$), and $L$ the membrane thickness (m). The surface porosity $\varepsilon$ of a separation membrane is the fraction of the surface area that is occupied by pore openings. As mentioned above, the tortuosity $\tau$ is the ratio of the mean length of the curvilinear transport pathways through the membrane ($L_s$) and *L*. If the BCP membranes contain straight cylindrical pores, the tortuosity $\tau$ equals 1. For all other pore morphologies, $\tau$ is larger than 1 because even the shortest curvilinear connection between the membrane surfaces is then longer than *L* (Fig. 1). Consequently, the effective pore diameter *d* of a membrane can be estimated either from the *MWCO* values or from $v$ – in addition to other texture characterization methods such as porosimetry. Especially when swollen hydrophilic test polymer molecules adsorb on membrane pore walls in the course of membrane separation tests, effective pore diameters *d* estimated from separation selectivities and $v$ may be smaller than the textural pore diameters of dry membranes measured by porosimetry. The pore geometry generally impacts the permeance of a membrane.[21] Thus, non-cylindrical cross sections of the membrane pores result in modified water flow. For example, cylindrical domains in BCP films oriented parallel to the film plane were converted into slit-shaped pores at the positions of the cylindrical domains.[9] The slit-shaped pores had widths of around 10 nm, but lengths of several µm up to tens of µm. For the description of water flow through the slit-shaped pores a modified Hagen-Poiseuille's law was proposed,[9, 21] where *h* denotes the width of the slit-shaped pores:

$$v = \frac{\varepsilon}{\tau^2}\left(\frac{\Delta p h^2}{3\mu L}\right) \quad (5)$$

**2.3 Membrane design: major challenges**

Although considerable progress has been made in the design and performance optimization of BCP membranes, several problems need to be addressed to push forward the use of BCP separation membranes in real-life applications. As already mentioned, the design of separation membranes generally involves a trade-off between separation selectivity, permeance and mechanical stability of the membranes that impedes performance breakthroughs.[22] High permeances are desired because high permeances result in higher water throughput. Thus, more water can be purified per time and membrane area. On the other hand, for obvious reasons also high separation selectivities are required. Pore length and membrane thickness influence the selectivity only to some extent if the membrane pores are large enough for the occurrence of convective fluid transport.[23] In the case of ultra- and nanofiltration through membrane pores with diameters well below 100 nm, exclusively the ratio between solute diameter and pore diameter should determine selectivity. Then, the selectivity can be optimized by carefully adjusting the diameter of the membrane pores and by the realization of narrow pore size distributions. However, any combination of geometric and morphological parameters, such as membrane thickness, pore diameter and porosity, is associated with drawbacks. Thicker separation membranes with smaller pore diameters will exhibit high separation selectivities but poor permeances. Increasing the pore diameter will result in better permeance but poorer selectivity. The membrane thickness is positively related to the transport resistance encountered by a fluid passing the membrane – the thinner a membrane is, the higher is its permeance. Therefore, thin membranes with small pores may show good permeances as well as good separation selectivities. Unfortunately, thin membranes typically lack sufficient mechanical robustness for routine operations. Reducing porosity while keeping pore diameter and membrane thickness constant improves mechanical stability but reduces in turn permeance – possibly below acceptable levels. As discussed below, only complex composite and gradient membrane configurations exhibit combinations of properties sufficient for real-life use cases. In addition, organic and biological fouling often occurs during the use of separation membranes. Membrane fouling related to the chemical and physical surface properties of the pore walls reduces permeance.[10] Contaminants with sizes comparable to the sizes of the membrane pores may get stuck in the membranes pores. The resulting pore clogging reduces the permeance. These problems may be tackled by sophisticatedly designed BCP-based separation membrane configurations, as, for example, comprehensively discussed by Hampu et al.[17]

# 3. BCP ultrafiltration membranes – a modular assembly system

## 3.1 BCPs as a material platform for ultrafiltration membranes

BCPs are macromolecules consisting of two or more chemically different block segments with distinct physicochemical properties. Typically, the different block segments of a BCP are immiscible and tend to segregate. However, since the block segments are connected by covalent bonds, macroscopic phase separation is impossible. Instead, microphase separation takes place and typically results in the formation of an ordered nanoscopic domain structure. The domains contain one specific block segment species; the covalent bonds between the different block segments enrich, therefore, at the domain interfaces. The morphology of the nanoscopic BCP domain structures is determined by parameters like the volume fractions occupied by the different BCP block segment species and their degree of compatibility. Symmetric diblock copolymers, in which the two components occupy half of the volume, form lamellar equilibrium morphologies. In asymmetric diblock copolymers, the equilibrium morphologies change from gyroid to cylindrical to spherical as the volume fraction of the minority component decreases.[24] Defects in the nanoscopic BCP domain structures may be healed by methods such as solvent annealing, thermal annealing and application of magnetic or electric fields.[15] The periods of the nanoscopic domain structures formed by BCP microphase separation typically amount to a few 10 nm. For a given nanoscopic domain structure type, the periods of the BCP domains may be tuned by the degree of polymerization of the BCP block segments (while their volume fractions are kept constant).

The chemical synthesis of tailored BCPs is well established and comprises synthetic strategies such as anionic polymerization, atom transfer radical polymerization and reversible addition-fragmentation chain transfer radical polymerization.[25] Hence, a modular assembly system of BCP architectures is accessible by state-of-the-art polymer chemistry. The chemical composition and the length of the BCP block segments as well as the sequence, in which they are incorporated into BCPs, can be customized. Customizing the chemical structure of BCPs in turn allows tailoring their morphology and period as well as chemical contrast and thermodynamic compatibility of their block segments. Besides diblock copolymers (AB) as the simplest BCP architecture, triblock copolymers (ABA, ABC), quarterpolymers (*e.g.*, ABAB or ABCD) as well as star-, comb- and cyclic BCPs can be synthesized. Hence, a wealth of different BCP architectures is available,[25] which have attracted tremendous interest in nanoscience – in particular for the preparation of nanoporous films, membranes and monoliths.

## 3.2 Textural design of BCP membranes

The pore systems of BCP membranes are typically derived from the nanoscopic BCP domain structures resulting from microphase separation in the BCP films and monoliths used as starting materials for the preparation of the BCP membranes. As discussed below, important textural features of BCP membranes, such as pore sizes, pore shapes and membrane porosities, are crucially influenced by the nature of the nanoscopic BCP domain structures that template pore formation at least to some extent.

### 3.2.1 Pore size

The possibility to customize pore sizes in membranes is an efficient lever for the tailoring of permeance and selectivity. BCP membrane pores derived from nanoscopic BCP domain structures typically have pore diameters corresponding to or being slightly larger than the domain sizes of the parent nanoscopic BCP domain structures. Therefore, the pores of BCP membranes typically have sizes from several nm up to several tens of nm, but well below 100 nm. Considering the pore size range of BCP membranes, BCPs are a promising material platform for the design and synthesis of ultrafiltration membranes and, to some extent, of nanofiltration membranes. The *MWCO* is an appropriate benchmark parameter for BCP membranes because the molecular dimensions of synthetic and biological polymers typically match their pore sizes.

The derivation of the pores in BCP membranes from regular nanoscopic BCP domain structures typically results in narrow pore size distributions. Pore size and pore size distribution of BCP membranes can be tuned and tailored much more efficiently than pore size and pore size distribution of homopolymer membranes, because classical pore formation techniques for the preparation of homopolymer membranes based on spinodal decomposition[26] are difficult to control. Narrow pore size distributions are a striking advantage of BCP membranes over most separation membranes consisting of other materials, because narrow pore size distributions boost selectivity. More specifically, narrow pore size distributions are the prerequisite for a sharp stepwise increase of the rejection rates for macromolecules if their molecular weights become equal or larger than the *MWCO* value. In contrast, separation membranes with broad pore size distributions will be characterized by a diffuse molecular weight range, in which the rejection rates for macromolecules gradually increase with the molecular weight, rather than by well-defined *MWCO* values.

### 3.2.2 Pore shape

Pore shapes crucially influence the separation performance of any kind of separation membrane. If a BCP membrane has an interpenetrating-bicontinuous morphology consisting of a continuous pore network and a likewise continuous scaffold, the length of the transport paths through the membrane exceeds the membrane thickness. Consequently, $\tau$ will be larger than 1, and the larger $\tau$ is, the longer are the transport paths. BCP separation membranes with spongy tortuous pore systems may show excellent selectivity but poor permeance, because pore walls intersecting straight transport paths are obstacles to fluid flow. Another drawback of spongy tortuous pore systems is their susceptibility to membrane fouling. Membrane fouling involves adsorption of larger macromolecular solutes or of particulate pollutants at the pore walls. As a result, the pore cross section is reduced, or the pores are even clogged. To optimize permeance, transport paths through BCP membranes should be as short as possible. It is straightforward that the length of the shortest possible transport path through a membrane corresponds to the membrane thickness $L$. Such short transport paths can be realized if the membranes are penetrated by straight pores having their pore axes oriented perpendicularly to the membrane plane – corresponding to a tortuosity of 1. Such membranes are referred to "homoporous".[13]

### 3.2.3 Porosity

The porosity is the volume fraction of a porous material that is occupied by the pores. High porosities result in high permeances but thinner pore walls. The latter feature may compromise the mechanical stability of a membrane so that it cannot withstand the impact that is exerted on it during use. Lowering the porosity increases the mechanical stability of a membrane but reduces permeance. The surface porosity $\varepsilon$, which is the area fraction of the pore openings at the surface of a porous material, may differ from the bulk porosity inside a membrane. A low surface porosity reduces the permeance of a membrane. In the case of homoporous BCP membranes containing arrays of straight cylindrical pores with constant diameter along their length, which are oriented normal to the membrane plane, bulk porosity and surface porosity have the same value – typically up to 15%.[20, 27] The porosities of BCP membranes with slit-shaped pores may be as high as 34 %.[9] Spongy interpenetrating-bicontinuous BCP membranes containing continuous pore networks may possess bulk porosities of up to 60 %.[13]

### 3.3 Generation of pores

A plethora of methods for the generation of pores in BCP films and monoliths has been established. The overarching design criterion for pore systems in BCP separation membranes is that the opposite membrane surfaces need to be connected by the pore systems. Moreover, the surface porosities need to be sufficiently large so that the pore systems can be accessed from the environment. Hence, pore systems of BCP separation membranes should provide transport pathways connecting the compartments separated by the BCP separation membranes.

Most pore generation routes yield nanosized pores in BCP membranes, which are suitable for ultrafiltration. In section 3.3 we will summarize basic pore formation strategies that allow the transformation of solid BCP films and monoliths into porous BCP films and monoliths. Pore formation processes coupled with the generation of complex membrane configurations to tackle the trade-off between permeance and selectivity, such as pore formation in the course of self-assembly combined with non-solvent induced phase separation (SNIPS),[28] are discussed in the following section 4.

**3.3.1 Selective degradation of block segments**
To generate pores in asymmetric microphase-separated BCPs, in which the block segments of a majority component form a continuous matrix, the minority domains may be degraded. The majority domains are conserved as walls of the pores formed in place of the minority domains.[12] For example, degradation of cylindrical minority domains yields holey films of the remaining majority component containing cylindrical pores. Several degradation methods including hydrolysis, UV irradiation, ozonolysis and treatment with acids such as hydrogen fluoride (HF), hydroiodic acid (HI) and trifluoroacetic acid (TFA) were employed to generate nanoporous BCPs. Selective degradation of one of the block segment species of a BCP by acidic or basic hydrolysis can be accomplished if at least another block segment species is inert under the applied conditions. Frequently, polyesters such as poly(lactide) (PLA) are employed as hydrolysable block segments of, for example, polystyrene-*block*-polylactide (PS-*b*-PLA).[29] Porous polymer membranes are also accessible by hydrolysing the poly(propylene carbonate) (PCC) block segments of poly(propylene carbonate)-*block*-poly(4-vinylcatechol acetonide) (PPC-*b*-PVCA) in alkaline solution.[7] The poly(methyl methacrylate) (PMMA) block segments of polystyrene-*block*-poly(methyl methacrylate) (PS-*b*-PMMA) selectively decompose by exposure to UV radiation; the short oligomeric PMMA chains thus formed can then be extracted with selective solvents such as acetic acid. Notably, the UV irradiation induces cross-linking of the PS chains in the remaining PS scaffold. The crosslinks between different PS chains improve the mechanical strength of the nanoporous PS membranes obtained in this way.[30] A drawback of pore formation by selective degradation of a BCP block segment is the mechanical weakening of the remaining membrane scaffold. The partial destruction of the BCP molecules often results in deteriorated mechanical properties of BCP separation membranes obtained in this way.

**3.3.2 Selective extraction of dissolvable species from minority domains**
As an alternative to their decomposition, the BCP block segments forming the minority domains can be loaded with a dissolvable species. The dissolvable species needs to be selectively compatible with the minority block segments so that it segregates into the minority domains during microphase separation. The dissolvable species is then extracted from the BCP minority domains with a selective solvent, against which the majority domains are inert. As a result, the nanoscopic domain structure of the BCP is arrested by the rigid majority component, while pores having walls consisting of the minority block segments form in place of the former hybrid domains containing the minority block segments and the dissolvable species. The dissolvable species may be a short homopolymer having the same chemical repeat unit than the BCP's minority block segments so that homopolymer and BCP minority block segments are miscible. In an early example, the PMMA minority domains in a film consisting of asymmetric PS-*b*-PMMA were blended with PMMA homopolymer. The film was formed on a modified substrate with balanced substrate-PS and substrate-PMMA interfacial interactions. Thus, hexagonal cylindrical domains containing the PMMA block segments and the PMMA homopolymer oriented normal to the film plane formed. Treatment with acetic acid removed the PMMA homopolymer so that the PMMA cylinders were converted into cylindrical pores with walls consisting of PMMA block segments.[31]

Alternatively, the block segments forming the minority domains can be loaded with small molecules. The miscibility of a polymer and a low-molecular-mass compound is typically better than the miscibility of two polymeric species, because in the former case a significant amount combinatorial entropy of mixing is generated, whereas in the latter case the generated combinatorial entropy of mixing is negligible. Commonly, small molecules are chosen that undergo specific noncovalent interactions with the BCP minority block segments, such as halogen bonding or hydrogen bonding. While these noncovalent interactions help direct the small molecules into the minority domains of the BCP, the small molecules can nevertheless be extracted by treatment with suitable polar solvents. As a result, again pores form that mimic the shape of the minority domains and that have pore walls consisting of the minority block segments. A typical BCP used in this context is asymmetric poly(styrene)-*b*-poly(4-vinylpyridine) (PS-*b*-P4VP) with PS as majority component forming a continuous matrix surrounding minority domains consisting of P4VP. Each P4VP repeat unit contains a pyridyl group with an N atom that can act as Lewis base. Small molecules with suitable functional groups such as hydroxyl groups or carboxyl groups may form hydrogen bonds with the N atom of P4VP.[32] A general limitation associated with pore formation by extraction of dissolvable species from BCP minority domains is that only a limited amount of the dissolvable species can be incorporated into the minority domains. If the amount of the dissolvable species is too high, the volume fraction of the minority domains may increase to such an extent that the type of the nanoscopic domain structure changes, *e.g.*, from cylindrical to gyroidal. If the content of the dissolvable component is further increased, macroscopic phase separation into a BCP-rich phase and a phase, in which the dissolvable species is enriched, may occur.

### 3.3.3 Selective swelling-induced pore generation

Selective swelling-induced pore formation[13, 33] is a simple process involving the treatment of a BCP film or a BCP monolith with a solvent selective to the minority domains. In the case of thin BCP films with a thickness comparable to the BCP period typically only surface reconstruction or formation of shallow concave indentations occurs.[34] For the production of BCP separation membranes predominantly swelling-induced pore formation in – often unsupported – BCP monoliths with thicknesses of a few times the BCP period is relevant.[13, 33] Starting point is the deposition of a BCP solution on a substrate yielding solid microphase-separated BCP films or monoliths. Exposure to a selective solvent for the minority block segments leads to selective migration of the solvent molecules into the minority domains. In analogy to osmosis, more and more solvent molecules are drawn into the minority domains that consequently swell. The minority block segments maximize contact with the solvent molecules by adapting stretched conformations. The majority component of the BCP also swells but to a much lower extent than the strongly swollen minority domains. The osmotic pressure emerging from the swollen minority domains triggers plastic deformation of the majority domains so as to accommodate the increasing volume of the minority domains. Any transient morphology can be conserved by quenching selective swelling-induced pore formation. For this purpose, the selective solvent is simply removed. Once the selective solvent evaporates, the stretched block segments of the BCP minority component undergo entropic relaxation into coiled conformations, while the majority component constitutes a rigid scaffold that fixates the reconstructed transient morphology. As a result, pores with walls consisting of the collapsed minority block segments form in place of the previously swollen minority domains. Swelling-induced pore formation can be applied to a wide range of BCPs consisting of block segments, for which selective solvents exist. In particular PS-*b*-PxVP (x=2, 4) has been used frequently to prepare nanoporous membranes by selective swelling-induced pore formation,[9, 33, 35] because PS and PxVP are highly incompatible so that highly selective solvents for both components exist.

Mild swelling strengths achievable with suitable solvents by appropriately adjusting swelling temperatures and durations yield pores faithfully mimicking the arrangement and the geometry of the minority domains prior to swelling. In this selective swelling-induced pore formation regime, the nanoscopic BCP domain structures template the formation of regular pore systems with narrow pore size distributions. The formed pore systems thus retain the initial ordering of the nanoporous domain structures. To some extent, pore sizes and membrane porosities can be tuned without deterioration of the regular pore arrangement and without significant broadening of the pore size distribution.[9, 35] The conservation of the ordering of the nanoscopic BCP domain structures combined with the textural design flexibility of selective swelling-induced pore formation is advantageous for the fabrication of BCP membranes with excellent selectivity and adjustable permeances. For example, selective swelling-induced pore generation of PS-*b*-P2VP films containing P2VP cylinders oriented parallel to the film surface yielded PS-*b*-P2VP membranes with highly regular, interconnected slit-shaped pores.[9]

At later stages of swelling-induced pore formation, the volume of initially spherical or cylindrical minority domains increases to such an extent that they are converted to interconnected continuous networks. Quenching selective swelling-induced pore formation at this stage yields spongy-continuous pore networks. Hence, selective swelling-induced pore formation is a viable access to mechanically stable BCP membranes with bicontinuous morphology characterized by high surface and bulk porosities up to 60 %.[13] Since the starting point is a microphase-separated BCP characterized by a regular nanoscopic domain structure, the pore sizes lie well below 100 nm. The natural endpoint of selective swelling-induced pore formation would be the complete dissolution of the BCP film or monolith into micelles or inverse micelles corresponding to the equilibrium structure the BCP forms in the applied solvent.

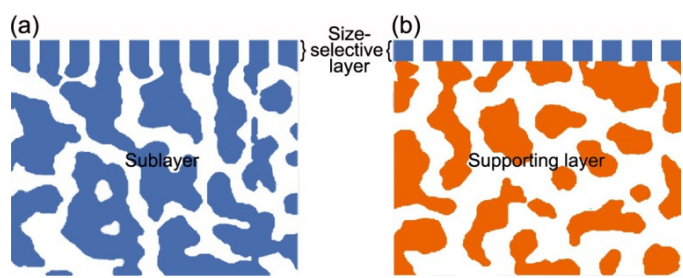

**Fig. 2.** Schematic illustration of the basic configurations of (a) monolithic BCP gradient membranes containing a thin size-selective layer on top of a macroporous sublayer and (b) composite membranes consisting of a thin size-selective BCP layer bonded on a macroporous supporting layer of a different material. BCPs are coloured blue, supporting layers of other materials are coloured orange.

## 4 Tackling the "trade-off" effect
### 4.1 General considerations
The optimization of the performance of separation membranes – including BCP-based ultrafiltration membranes – needs to address the competition between selectivity and permeance ("trade-off" effect). Thin homoporous BCP membranes with narrow pore size distributions may solve this problem. Such homoporous membranes may be derived from BCP films or monoliths containing arrays of cylindrical minority domains oriented normal or parallel to the film/monolith plane. For instance, 42 nm thick BCP membranes with slit-shaped pores exhibited a 10-100 times higher permeance than commercial membranes with comparable selectivity.[9] However, only few examples of thin unsupported BCP membranes with sufficient mechanical stability are known, such as porous polysulfone-*b*-poly(ethylene glycol) (PSF-*b*-PEG) membranes, in which the PSF domains exhibit excellent mechanical, thermal, and chemical stability.[36] Most thin BCP membranes lack the mechanical robustness required for use in real-life applications. The solution to this problem is the combination of thin homoporous BCP membranes exhibiting both high selectivity and high permeance with a supporting layer for mechanical stabilization. Two design options exist:

1) monolithic BCP gradient membranes (Fig. 2a) and 2) composite membranes (Fig. 2b). BCP gradient membranes are monolithic in that they consist of a single continuous BCP scaffold. However, by suitable preparation methods, a thin size-selective BCP skin layer with small pores can be generated on top of a macroporous BCP sublayer (Fig. 2a). The second option are composite membranes consisting of a thin size-selective porous BCP layer adhesively bonded to a macroporous supporting layer of a different material (Fig. 2b). The sublayer of monolithic BCP gradient membranes and the supporting layer of composite membranes mechanically support the thin size-selective BCP layers. However, their own pores are so large – typically several microns to several tens of microns – that they only slightly increase flow resistance and reduce permeance.

**4.2 Monolithic BCP gradient membranes by self-assembly combined with non-solvent induced phase separation (SNIPS)**

Gradient membranes prepared by the SNIPS process[28, 37] consist of a thin size-selective BCP layer containing nanopores monolithically joined to a sublayer of the same BCP with much larger pores. The size-selective layer with a thickness typically ranging from 200 to 300 nm contains cylindrical through-pores with diameters of a few 10 nm oriented normal to the layer plane.[28] The sublayer with a typical thickness of hundreds of µm contains a spongy pore system with low flow resistance. The SNIPS process involves simultaneous self-assembly and non-solvent induced phase separation of BCPs, which can be achieved in a technically simple way by casting a BCP film onto a substrate followed by immersion into a nonsolvent for all BCP components. A BCP frequently used to prepare SNIPS membranes is PS-*b*-P4VP. Tetrahydrofuran (THF) is then employed as a selective solvent for PS and dimethyl formamide (DMF) as a selective solvent for P4VP.[28] After casting a solution of PS-*b*-P4VP in a THF-DMF mixture onto a substrate, predominantly THF evaporates since THF is more volatile than DMF. THF thus depletes within tens of seconds at the solution-air interface, whereas the DMF concentration increases. The DMF selectively migrates into the P4VP-rich regions, which consequently swell, whereas the PS-rich regions shrink. These effects induce the alignment of cylindrical domains of the P4VP-rich phase normal to the film plane within a skin layer at the air surface of the PS-*b*-P4VP film. Subsequently, the PS-*b*-P4VP film is immersed into water, which is a nonsolvent for PS and P4VP. As a result, the swollen P4VP block segments in the skin layer collapse, and the P4VP cylinders normal to the film surface are converted into cylindrical pores with P4VP-lined pore walls. Away from the surface, a PS-*b*-P4VP-rich phase and a water-rich phase form by macroscopic phase separation. The PS-*b*-P4VP-rich phase vitrifies and forms the scaffold of the sublayer. After the removal of the water, the volume initially occupied by the water-rich phase forms macropores.

**4.3 Monolithic gradient membranes by selective swelling-induced pore generation**

Monolithic gradient membranes are also accessible by a two-step selective swelling-induced pore formation procedure.[35] A first, mild selective swelling-induced pore formation steps convert the minority domains of a given nanoscopic BCP domain structure generated by microphase separation into pores. The formed pores morphologically mimic the minority domains so that the initial ordering of the minority domains is retained. If, for example, PS-*b*-P2VP ($M_n$(PS) = 50 kg/mol; $M_n$(P2VP)=16.5 kg/mol) is subjected to a first swelling step involving treatment with ethanol at 60°C for 1 h, pores with a diameter of 8 nm are obtained, and a permeance of 83 L·m$^{-2}$·h$^{-1}$·bar$^{-1}$ is achieved. Subsequent irradiation of the PS-*b*-P2VP membrane with UV light results in cross-linking of the PS-*b*-P2VP molecules located close to the exposed membrane surface. The crosslinking irreversibly fixates the pore morphology formed by the initial mild selective swelling-induced pore formation step in a skin layer at the membrane surface. Finally, a second selective swelling-induced pore formation step is carried out under harsher conditions resulting in the formation of larger pores – except in the skin layer. The skin layer with a thickness of less than 100 nm serves as the size-selective layer of the monolithic gradient membrane obtained in this way, whereas the underlying sublayer with much larger pores formed by the second selective swelling-induced pore formation step mechanically stabilises the monolithic gradient membrane. The gradient membranes exhibit reduced mass-transfer resistance and thus an increased permeance of 295 L·m$^{-2}$·h$^{-1}$·bar$^{-1}$. However, the rejection rates after the first and second selective swelling-induced pore formation steps were similar and amounted to approximately 95 % for bull serum albumin (BSA, $M_w$ = 67 kDa, size: 14 x 4 x 4 nm$^3$). If the second selective swelling-induced pore formation step is carried out under even harsher conditions, for instance, at 60 °C for 15 h (after a first selective swelling-induced pore formation step at 50°C for 5 h), the permeance is increased to 557 L·m$^{-2}$·h$^{-1}$·bar$^{-1}$ while the BSA rejection still remained as high as 93%. Apparently, the selectivity of the cross-linked size-selective skin layer remains unaffected by the second selective swelling-induced pore formation step.

**4.4 Composite membranes**

The main advantage of composite membranes is that two or more functional layers with complementary properties and/or functions can be combined. The main lever to maximize permeance and selectivity of composite membranes is the minimization of the thickness of size-selective BCP separation layers characterized by narrow pore size distributions. The size-selective BCP separation layers are joined to mechanically stable macroporous supporting layers with large pore sizes, large porosities and high permeances. Typical supporting layers, which are either commercially available or accessible by well-established manufacturing processes, include monolithic macroporous polymer membranes consisting, for example, of PES or PVDF as well as nonwoven fabrics such as electrospun fiber mats. The BCPs forming the size-selective separation layers can be deposited onto the macroporous supporting layers by methods such as machine casting,[38] spin coating[8, 39], spray coating[40] or film transfer.[41, 42] Frequently, the size-selective BCP separation layers are derived from BCPs containing PS block segments. The pores in the size-selective BCP separation layers can be generated by any of the methods reviewed in Section 3.3. For example, size-selective BCP layers can be prepared by selective decomposition of the PLA block segments of poly(styrene)-*b*-poly(isoprene)-*b*-poly(lactide) (PS-*b*-PI-*b*-PLA) using bases so that pores form *in lieu* of the PLA domains.[8] The rubbery PI block segments improve the mechanical properties of the remaining nanoporous size-selective PS-*b*-PI layers. Another example involves the deposition of PLA-*b*-P(S-*s*-GMA) onto macroporous PES layers.[43] P(S-*s*-GMA), which is the statistical copolymer of styrene (S) and glycidyl methacrylate (GMA), was crosslinked, whereas the PLA block segments were degraded. In this way, composite membranes containing 500 nm thick size-selective P(S-*s*-GMA) layers with bicontinuous morphology and an average pore radius of 5 nm were obtained. Moreover, composite membranes containing size-selective nanoporous PVCA separation layers were obtained by alkaline hydrolysis of the PCC block segments of PPC-*b*-PVCA layers.[7] Size-selective BCP separation layers of composite membranes were also obtained by extraction of a dissolvable component mixed into BCP minority domains. Examples include the dissolution of PMMA homopolymer mixed with PMMA minority domains of PS-*b*-PMMA[31] as well as the extraction of small molecules mixed into the P4VP minority domains of PS-*b*-P4VP layers.[44] Composite membranes were also prepared by selective swelling-induced pore generation. 42 nm thick PS-*b*-P2VP films containing arrays of P2VP cylinders oriented parallel to the film plane were deposited onto macroporous PES supporting layers[9] and subsequently treated with ethanol at 50 °C for different durations to convert the P2VP cylinders into nanopores. Another example involves the transfer of 260 nm thick PS-*b*-P2VP films formed on water surfaces onto macroporous PVDF supporting layers followed by selective swelling-induced pore formation.[41]

Although selectivity and permeance of composite membranes depend by and large on the nature of the size-selective BCP membrane layers, the underlying macroporous supporting layers may also affect the permeance.[8] The pore sizes of the macroporous supporting layers should be as large as possible so that the flow resistance is as weak as possible. Therefore, the difference between the pore sizes of the size-selective BCP separation layer and the macroporous supporting layer should be as large as possible. However, if the pore size of the macroporous supporting layer is too large so that too large portions of the size-selective BCP separation are unsupported, the size-selective BCP separation layer may partially collapse. Moreover, a certain portion of the pore openings of the size-selective BCP separation layers may impinge on the pore walls of the macroporous supporting layer. A significant number of such dead-end pores may result in reduced permeance.[8] To circumvent this problem, macroporous supporting layers with ultrahigh surface porosities having pore walls formed by sharp scaffolds are used. Finally, sufficient adhesion between the size-selective BCP separation layers and the macroporous supporting layers is required to avoid delamination. Thus, besides the careful design of the size-selective BCP separation layers, the supporting layers must be carefully chosen also.

## 5. Applications of BCP ultrafiltration membranes

A large body of literature deals with the use of BCP separation membranes for ultrafiltration. In the following, some representative examples are discussed. The ultrafiltration performance of gradient membranes prepared by the SNIPS process was investigated intensively. For example, SNIPS membranes consisting of polystyrene-*block*-poly(2-hydroxyethyl methacrylate-*stat*-2-(succinyloxy)ethyl methacrylate) (PS-*b*-P(HEMA-*s*-SEMA)) contain pores lined with highly hydrophilic P(HEMA-*s*-SEMA) block segments characterized by abundant –OH and –COOH groups.[45] Aqueous solutions swell the P(HEMA-*s*-SEMA) block segments so that the pore diameter is markedly reduced. As a result, BSA permeates monolithic PS-*b*-P(HEMA-*s*-SEMA) gradient membranes, whereas haemoglobin (size: 7 x 5.5 x 5.5 nm$^3$) is rejected. In another example, poly-(2-(dimethylamino)ethyl methacrylate) (PDMAEMA) chains were grafted on graphene oxide sheets, which were in turn mixed with PS-*b*-P4VP. The PDMAEMA modification hydrophilized the graphene oxide sheets and prevented their agglomeration. Mixtures of PS-*b*-P4VP and of up to 1.5 wt-% PDMAEMA-modified graphene oxide were used to prepare PS-*b*-P4VP/graphene oxide hybrid gradient membranes with 200 nm thick size-selective layers containing straight aligned pores normal to the surface with an average diameter of 25 nm. The PS-*b*-P4VP/graphene oxide hybrid gradient membranes showed 1.6 times higher permeance and enhanced anti-fouling properties as compared to the corresponding pure PS-*b*-P4VP SNIPS membranes, while BSA could efficiently be separated from globulin-γ (IgG, $M_w$ = 150 kDa).[46] Deposition of polydopamine/cysteine onto SNIPS membranes consisting of PS-*b*-P4VP improved the membrane hydrophilicity and antifouling performance as well.[47]

Ultrafiltration using BCP-based composite membranes was intensively studied also. For example, P(S-*s*-GMA)]/PES composite membranes with 500 nm thick size-selective P(S-*s*-GMA)] layers had a water permeance of 7 L·m$^{-2}$·h$^{-1}$·bar$^{-1}$, which is much lower than that of the bare PES supporting layer (3700 L·m$^{-2}$·h$^{-1}$·bar$^{-1}$). However, the rejection of fluorescent TRITC-dextran (0.5 mg·mL$^{-1}$, M = 150 kg·mol$^{-1}$, hydrodynamic radius ($r$) = 7 nm) was determined to be 98% by UV–vis absorbance spectroscopy. Decreasing the thickness of the size-selective BCP layer to 150 nm resulted in an increased permeance of 196 L·m$^{-2}$·h$^{-1}$·bar$^{-1}$, while the rejection of TRITC-dextran dropped to 96% as a signature of the trade-off effect.[43] Other examples are composite membranes containing size-selective PVCA separation layers obtained with PPC-*b*-PVCA as starting material. The pore walls of the size-selective PVCA separation layers were hydrophilized by ZrO$_2$ coatings generated by a biomineralization process.[7] For this purpose, the catechol acetonide side groups of the PVCA chains were converted to catechol side groups, which in turn chelated Zr$^{4+}$ supplied as acidic zirconium sulfate solution. Thus-obtained composite membranes containing 200 nm thick size-selective PVCA-ZrO$_2$ separation layers with a pore diameter of 12 nm had a permeance as high as 114 ± 4 L·m$^{-2}$·h$^{-1}$·bar$^{-1}$. The permeance of ZrO$_2$-functionalized membranes was about 5 times higher than that of composite membranes with untreated size-selective PVCA layers. The retention capability of composite membranes containing size-selective PVCA-ZrO$_2$ separation layers was demonstrated by completely removing Au nanoparticles with a diameter of 20 nm from aqueous dispersions, and their antifouling properties were demonstrated by protein adsorption experiments.

The permeance of composite membranes consisting of 260 nm thick size-selective PS-*b*-P2VP separation layers prepared by selective-swelling-induced pore generation in ethanol at 55 °C for 1 h and of macroporous PVDF supporting layers amounted to 319 L·m$^{-2}$·h$^{-1}$·bar$^{-1}$.[41] Extending the ethanol treatment to 3 h resulted in an increased permeance of 832 L·m$^{-2}$·h$^{-1}$·bar$^{-1}$ attributed to the formation of larger pores. The rejection of BSA in turn declined from 93 % to 81 %. Further extension of the exposure to ethanol to 12 h resulted in a significantly enhanced permeance of 1080 L·m$^{-2}$·h$^{-1}$·bar$^{-1}$, while the rejection of BSA fell to 50%. The performance of composite membranes can be optimized by modifications of size-selective BCP layers prepared by selective swelling-induced pore formation. For example, the hydrophilicity of nanoporous PS-*b*-PDMAEMA membranes obtained by selective swelling-induced pore formation was increased by functionalization of the pore walls with zwitterions, resulting in several times higher permeances.[48]

| Table 1. Advantages of BCP membranes for water purification | | | | | |
|---|---|---|---|---|---|
| **Advantage** | High selectivity | High permeance | Good hydrophilicity and antifouling | Outstanding design flexibility regarding morphology and chemistry | Device integration |
| **Responsible design feature** | Narrow pore size distribution in size-selective layers | - Ultrathin size-selective layers<br>- straight pores normal to membrane plane | Hydrophilic pore walls | Polymer chemistry as versatile modular assembly system for BCPs | Integration of size-selective BCP layers in composite and gradient membranes |
| **Examples** | ref. 31 | ref. 39 | ref. 39, 47 | ref. 7, 50 | ref. 8, 37 |

## 6. BCP membranes for nanofiltration

The periods of the nanoscopic BCP domain structures, which template the pore formation in the course of the preparation of BCP membranes, typically amount to a few 10 nm. Therefore, the direct preparation of nanofiltration membranes with effective pore diameters of 2 nm or smaller using BCPs as starting materials is challenging. Instead of applying one of the classical mechanisms for pore formation in BCPs reviewed in Section 3.3, Tu *et al.*[49] introduced β-barrel membrane proteins (MPs) into PB-*b*-PEO (PB = polybutadiene) bilayers by a solution-based self-assembly process. In this way, PB-*b*-PEO nanosheets containing MPs providing transport channels normal to the nanosheet plane with effective sizes of 0.8, 1.3 or 1.5 nm (depending on the MP used) were obtained. To integrate the nanosheets into continuous size-selective separation layers, the classical layer-by-layer technique for the alternating deposition of oppositely charged polyelectrolytes was adapted as follows. Positively charged polyethylenimine (PEI) was deposited onto a negatively charged macroporous support layer. The nanosheets were negatively charged by functionalisation of the PB-*b*-PEO with carboxylic acid groups. These modified, negatively charged nanosheets were deposited onto the positively charged PEI layer. The successive deposition of oppositely charged PEI and modified nanosheet layers was repeated a few times, and the obtained membranes were additionally stabilized by chemical crosslinking. The *MWCO* value of the PB-*b*-PEO/MP hybrid membranes obtained in this way was as small as 477 Da, while a water permeance of 293 ± 51 L·m$^{-2}$·h$^{-1}$·bar$^{-1}$ was obtained. PB-*b*-PEO/MP hybrid membranes with an increased water permeance of 1092 ± 79 L·m$^{-2}$·h$^{-1}$·bar$^{-1}$ still had a *MWCO* value of 930 Da. The water permeance of PB-*b*-PEO/MP hybrid membranes was up to three orders of magnitude higher than that of commercial reference membranes with comparable selectivity. However, the complex fabrication of PB-*b*-PEO/MP hybrid membranes may impede their wide use.

Size-selective nanofiltration layers were also directly derived from BCPs. For example, the pores of size-selective BCP separation layers can be physically narrowed by selective swelling of the BCP block segments lining the pores under operating conditions. In contrast to selective swelling-induced pore formation (Section 3.3.3), the extent of swelling must be moderate enough so as not to alter the textural structure of the BCP separation layer. However, crosslinking the block segments constituting the scaffolds of the size-selective BCP separation layers may significantly enhance the stability of the latter during operation. Also, electrostatic repulsion exerted by charged pore walls can be exploited to achieve nanofiltration, provided the solutes carry the same charges than the pore walls. For example, the pore walls of monolithic P(HTMB-*r*-I)-*b*-PS-*b*-P4VP gradient membranes prepared by the SNIPS process (Section 4.2) were lined with the two miscible end blocks P(HTMB-*r*-I) and P4VP.[37] Treatment with methyl iodide yielded positively charged, treatment with 1,3-propane sultone negatively charged P4VP blocks at the pore walls. While in this way the pore diameters in the dry state were only slightly reduced from 32 nm to 30 nm, the permeance for ultrapure water with a pH value of 5.5 was reduced from 515 $L·m^{-2}·h^{-1}·bar^{-1}$ for as-prepared P(HTMB-*r*-I)-*b*-PS-*b*-P4VP gradient membranes to 11 and 9.5 $L·m^{-2}·h^{-1}·bar^{-1}$ for P(HTMB-*r*-I)-*b*-PS-*b*-P4VP gradient membranes treated with methyl iodide and 1,3-propane sultone. The ionized polyelectrolytic P4VP chains were swollen during water filtration so that the effective pore sizes were sharply reduced. Charge-based separation in the nanofiltration regime was apparent from the rejection of 95.3% of the cationic dye methylene blue (size: 1.1 nm) by P(HTMB-*r*-I)-*b*-PS-*b*-P4VP gradient membranes treated with methyl iodide. P(HTMB-*r*-I)-*b*-PS-*b*-P4VP gradient membranes treated with 1,3-propane sultone rejected 95.2 % of trivalent Naphthol Green B (diameter 1.8 nm). Also, post-functionalization of PS-*b*-P(HTMB-*r*-I) [50] and PS-*b*-P4VP[51] gradient membranes prepared by the SNIPS process resulted in the rejection of dyes with a size of less than 2 nm following the same principle.

A slightly different modification strategy was applied in the case of polyisoprene-*b*-polystyrene-*b*-poly(N,N-dimethylacrylamide) (PI–*b*-PS-*b*-PDMA) gradient membranes obtained by the SNIPS process.[52] The side groups of the PDMA block segments were selectively hydrolysed. In this way, the amide groups were converted into carboxyl groups so that the membrane pores were lined with PAA (polyacrylic acid) block segments. The PAA segments were then converted into poly(2-acrylamido-ethane-1,1-disfulonic acid) (PADSA) block segments by a carbodiimide coupling reaction, which enables the attachment of sulfonic acid moieties onto the pore walls. The reduction of the effective pore diameter to 2 nm for the PI–*b*-PS-*b*-PADSA membranes resulted in the rejection of 94 % of oligomeric ethylene oxide with a hydrodynamic radius below 1 nm, demonstrating robust size-based selective nanofiltration performance.

Most examples of BCP-based nanofiltration membranes are based on complex configurations of the size-selective separation layers or rely on specifically synthesized BCPs that are either not commercially available or rather expensive. In order to pave the way for the upscaling of the production of BCP-based nanofiltration membranes, the commercially available BCP poly(styrene-*b*-butadiene-*b*-styrene) (PS-*b*-PB-*b*-PS) was used as starting material. PS-*b*-PB-*b*-PS solutions were cast on polyacrylonitrile substrates to prepare gradient membranes.[53] The hydrophobic nature of the PS and PB block segments impedes the use of PS-*b*-PB-*b*-PS membranes for separation problems involving aqueous solutions. To overcome this drawback, a photomodification route using a thio-ene reaction was employed to introduce carboxylic groups into the backbones of the PB block segments. The membrane pores were thus covered with hydrophilized PB chains and the modified size-selective PS-*b*-PB-*b*-PS separation layers enabled efficient water purification. Although the membranes had an effective pore size of 4.4 nm based on the estimation of PEO rejections, the rejection rates of methyl orange (molecular mass is 327 $g·mol^{-1}$) and brilliant blue (molecular mass is 826 $g·mol^{-1}$) amounted to up to 74 % and 100 %, respectively. The nanofiltration capability of the modified PS-*b*-PB-*b*-PS membranes relies on electrostatic repulsion rather than on size-based separation as both the carboxyl groups of the modified PB block segments and the two selected model dyes are negatively charged.

## 7 Conclusions and Perspectives

Removing hazardous nanoparticulate contaminants, biological or synthetic macromolecules and viruses from fluids such as water and air is a technical challenge, the importance of which cannot be underestimated. Corresponding purification processes are crucial for the sufficient supply of clean water for a growing world population, but also for the containment of diseases spreading by transmission of viruses. Separation membranes for water and air purification should be characterized by high separation selectivity and high permeance. However, the design of separation membranes typically involves a trade-off between permeance and selectivity. Structural features favouring high permeance such as large pore sizes reduce selectivity. Small pore sizes ensure high selectivity but reduce permeance. The use of BCPs as a material platform for the manufacturing of separation membranes may reconcile high permeance with high selectivity because BCPs combine the high manufacturing flexibility of polymers with the possibility to generate tailored pore systems. The pore systems in BCP separation membranes are derived from the ordered nanoscopic domain structures the BCPs form by microphase separation. Conversion of the ordered nanoscopic BCP domain structures into pores may accomplished by selective degradation of block segments forming the BCP minority domains, by extraction of dissolvable species from BCP minority domains or by swelling-induced pore formation. Membrane configurations combining high permeance and high selectivity contain thin size-selective BCP separation layers with a thickness not exceeding a few 100 nm, with pores having a tailored diameter in the sub-100-nm range and with a sharp pore size distribution. Free-standing and unsupported thin size-selective BCP separation layers with sufficiently high porosity are mechanically not robust enough under operating conditions. Therefore, thin size-selective BCP separation layers are typically integrated into monolithic gradient membrane configurations or into composite membrane architectures. In both cases, macroporous layers with high permeance mechanically support the thin size-selective BCP separation layers. The advantages of BCP membranes are summarized in Table 1.

BCP-based monolithic gradient membranes and composite membranes with thin size-selective BCP separation layers show promising separation performances in the ultrafiltration regime, while their applicability in the nanofiltration regime is limited. However, so far, the investigation of BCP-based separation membranes has predominantly been limited to well-designed test scenarios using selected test solutes in laboratory environments. In contrast, real-life application scenarios have rarely been studied. One example involves the use of PSF-*b*-PEG ultrafiltration membranes for hemodialysis,[54] suggesting that BCP-based membrane systems may by applied in high-added-value fields such as health care. Until now, the manufacturing costs of BCP-based separation membranes are relatively high because BCPs as starting materials are relatively expensive. The production of affordable BCP-based separation membranes is still a challenge. Moreover, most of the BCPs currently used for the preparation of separation membranes contain PS block segments that form the membrane scaffold. Hence, the glass transition temperature of PS (~100 °C for bulk PS) limits the temperature range within which PS-containing BCP separation membranes can be used. If purification processes are carried out at elevated temperatures, BCP-based separation membranes may lack the required thermal stability. BCPs containing PSF block segments with a glass transition temperature of nearly 190 °C as scaffold-forming component may mitigate this drawback. PSF-*b*-PEG as starting material for membrane production may yield affordable and mechanically robust BCP membranes.[36] However, further efforts are required to narrow the pore size distribution of PSF-*b*-PEG separation membranes to improve their separation selectivity.[36] Finally, it should be noted that most BCP-based separation membrane configurations are produced in the form of flat sheets. It would be rewarding to integrate BCP separation layers into other geometries such as hollow-fibre membranes that may allow optimization of throughput and better device integration.[15]

**Conflicts of interest**
There are no conflicts to declare.

**Acknowledgements**
L. G. and M. S. acknowledge support by the European Research Council (ERC-CoG-2014, Project 646742 INCANA). Y. W. thanks financial support by National Science Fund for Distinguished Young Scholars (21825803).**Notes and references**